\documentclass[12pt]{article}
\usepackage{graphicx}
\usepackage{cite}
\usepackage{tabularray}
\usepackage{xcolor}

\newcommand\pubnumber{PSI-PR-23-45, ICPP-75}

\newcommand\pubdate{\today}

\def\instituteone{Physik-Institut, Universität Zürich, Winterthurerstrasse 190, CH–8057 Zürich, Switzerland}
\def\institutetwo{Paul Scherrer Institut, CH–5232 Villigen PSI, Switzerland}
\def\institutethree{School of Physics and Institute for Collider Particle Physics, University of the Witwatersrand,
Johannesburg, Wits 2050, South Africa}
\def\institutefour{iThemba LABS, National Research Foundation, PO Box 722, Somerset West 7129, South Africa}

\def\authemailsumit{\footnote{sumit.banik@physik.uzh.ch ($^*$speaker)}}
\def\authemailgug{\footnote{guglielmo.coloretti@physik.uzh.ch}}
\def\authemailandi{\footnote{andreas.crivellin@cern.ch}}
\def\authemailbruce{\footnote{bmellado@mail.cern.ch}}

\def\Title#1{\begin{center} {\Large #1 } \end{center}}
\def\Author#1{\begin{center}{ \sc #1} \end{center}}
\def\Address#1{\begin{center}{ \it #1} \end{center}}

\newcommand\pubblock{\rightline{\begin{tabular}{l} \pubnumber\\
         \pubdate  \end{tabular}}}
\newenvironment{Abstract}{\begin{quotation}  }{\end{quotation}}
\newenvironment{Presented}{\begin{quotation} \begin{center} 
             PRESENTED AT\end{center}\bigskip 
      \begin{center}\begin{large}}{\end{large}\end{center} \end{quotation}}
\def\Acknowledgements{\bigskip  \bigskip \begin{center} \begin{large}
             \bf ACKNOWLEDGEMENTS \end{large}\end{center}}

\def\beq{\begin{equation}}
\def\eeq#1{\label{#1}\end{equation}}
\def\eeqn{\end{equation}}

\def\beqa{\begin{eqnarray}}
\def\eeqa#1{\label{#1}\end{eqnarray}}
\def\eeqan{\end{eqnarray}}

\let\bar=\overbar

\def\Dslash{\not{\hbox{\kern-4pt $D$}}}
\def\dslash{\not{\hbox{\kern-2pt $\del$}}}

\def\msb{{\bar{\ssstyle M \kern -1pt S}}}

\begin{document}
\begin{titlepage}
\pubblock

\vfill
\Title{Differential $e\mu bb $ Cross-Sections and New Higgses at the Electroweak Scale}
\vfill
\Author{ Sumit Banik $^*$ \authemailsumit, Guglielmo Coloretti\authemailgug, Andreas Crivellin\authemailandi}
\Address{\instituteone \\[0.2cm] \institutetwo}
\Author{ Bruce Mellado\authemailbruce}
\Address{\institutethree \\[0.2cm] \institutefour}
\vfill
\begin{Abstract}
ATLAS found that none of their Standard Model simulations can describe the measured differential lepton distributions in their $t \bar{t}$ analysis reasonably well. Therefore, we study the possibility that this measurement has a new physics contamination. We consider a benchmark model motivated by the indications for di-photon resonances: A heavy scalar decays into two lighter Higgs bosons with masses of 152\,GeV and 95\,GeV, with subsequent decay to $WW$ and $bb$, respectively. In this setup, the description of data is improved by at least $5.6 \sigma$.
\end{Abstract}
\vfill
\begin{Presented}
$16^\mathrm{th}$ International Workshop on Top Quark Physics\\
(Top2023), 24--29 September, 2023
\end{Presented}
\vfill
\end{titlepage}
\def\thefootnote{\fnsymbol{footnote}}
\setcounter{footnote}{0}

\section{Introduction}

With the discovery of the Brout-Englert-Higgs boson~\cite{Higgs:1964ia,Englert:1964et,Higgs:1964pj,Guralnik:1964eu} at the Large Hardon Collider (LHC) in 2012~\cite{Aad:2012tfa,Chatrchyan:2012ufa}, the Standard Model (SM) has been completed. However, the minimality of the SM scalar sector is not guaranteed by any symmetry principle and a plethora of extensions, including the addition of $SU(2)_L$ singlets~\cite{Silveira:1985rk,Pietroni:1992in,McDonald:1993ex}, doublets~\cite{Lee:1973iz,Haber:1984rc,Kim:1986ax,Peccei:1977hh,Turok:1990zg} and triplets~\cite{Konetschny:1977bn,Cheng:1980qt,Lazarides:1980nt,Schechter:1980gr,Magg:1980ut,Mohapatra:1980yp} have been proposed.

In fact, there are indications of new Higgses at the electroweak (EW) scale with masses of $\approx95\,$GeV~\cite{LEPWorkingGroupforHiggsbosonsearches:2003ing,CMS:2018cyk,CMS:2022rbd,CMS:2022tgk,ATLAS:2023jzc} and $\approx152\,$GeV~\cite{ATLAS:2021jbf} with global significances of $3.8\sigma$ and $4.9\sigma$~\cite{Crivellin:2021ubm,Bhattacharya:2023lmu}, respectively. Furthermore, the ``multi-lepton anomalies''~\cite{Buddenbrock:2019tua,vonBuddenbrock:2020ter,Hernandez:2019geu,Fischer:2021sqw}, processes involving multiple leptons {and missing energy}, with and without (b-)jets, where significant deviations from the SM expectations have been observed over the last years, including $WW$, $WWW$, $Wh$, $tW$, $t\bar t$, $t\bar tW$ and $t\bar t t\bar t$ signatures (see Ref.~\cite{Fischer:2021sqw} and references therein) are compatible with, or even suggest, a new scalar with a mass of $\approx\!150\,$GeV~\cite{vonBuddenbrock:2017gvy,Coloretti:2023wng}.

In these proceedings, we will summarize the statistically most significant multi-lepton excess encoded in the latest ATLAS analysis~\cite{ATLAS:2023gsl} of the $t\bar t$ differential cross-sections as analyzed in Ref.~\cite{Banik:2023vxa}. We focus on the invariant di-lepton mass ($m^{e\mu}$) and the angle between the leptons ($\Delta \phi^{e\mu}$) for various combinations of Monte Carlo (MC) simulators.\footnote{Note that the CMS analysis of $t+W$~\cite{CMS:2018amb} is less precise but consistent with the ATLAS findings~\cite{Buddenbrock:2019tua}.} For this, we consider a simplified model in which a new scalar $H$ is produced via gluon fusion and decays to $S$ and $S^\prime$, with masses of 152\,GeV and 95\,GeV, respectively, as suggested by the hints for narrow resonances.

\begin{figure*}[h]
    \includegraphics[width=0.78\linewidth]{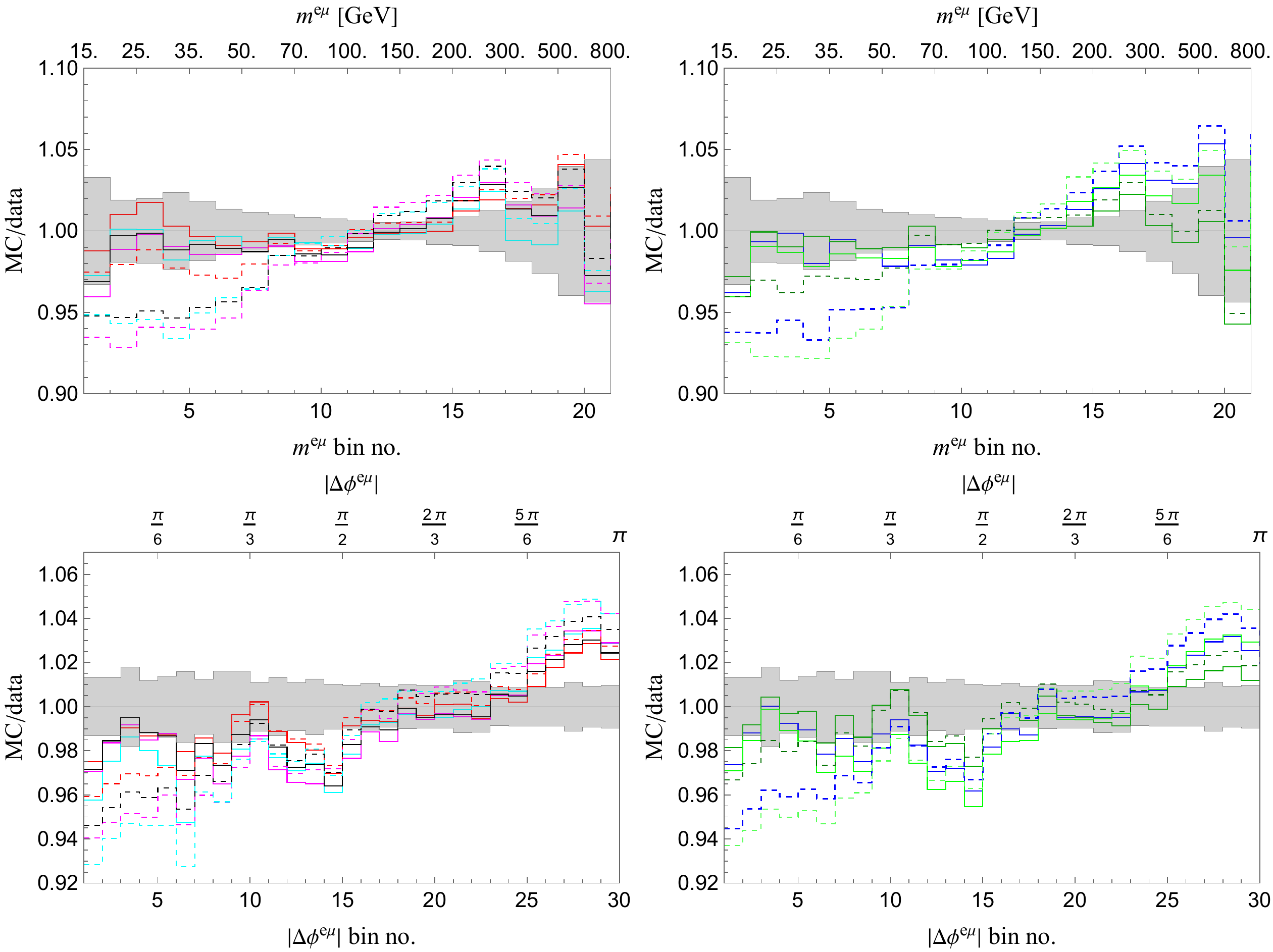}
    \raisebox{0.26\height}{\includegraphics[width=0.21\linewidth]{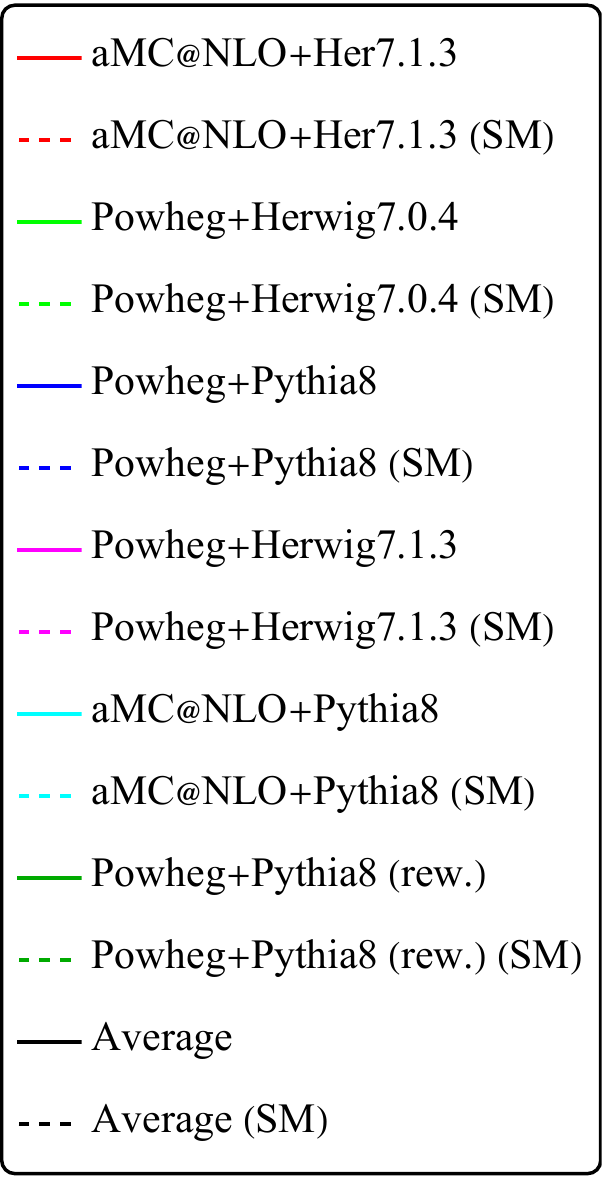}}
    \caption{The dashed coloured lines show six different SM predictions (MC) normalized to data given in Ref.~\cite{ATLAS:2023gsl}. The solid lines include the NP contribution from our benchmark model obtained by a combined global fit to $m^{e\mu}$ and $\Delta\phi^{e\mu}$ data including its correlations. The black lines are obtained by averaging the six predictions and the grey band shows the total uncertainty (systematic and statistical). One can see that the agreement between theory and experiment is significantly increased over the whole range of the distributions by adding an NP effect.}
    \label{fig:mlldata}
\end{figure*}

\section{Analysis and Results}
\label{sec:simplified}

\begin{figure}[h]
    \centering
    \includegraphics[scale=1]{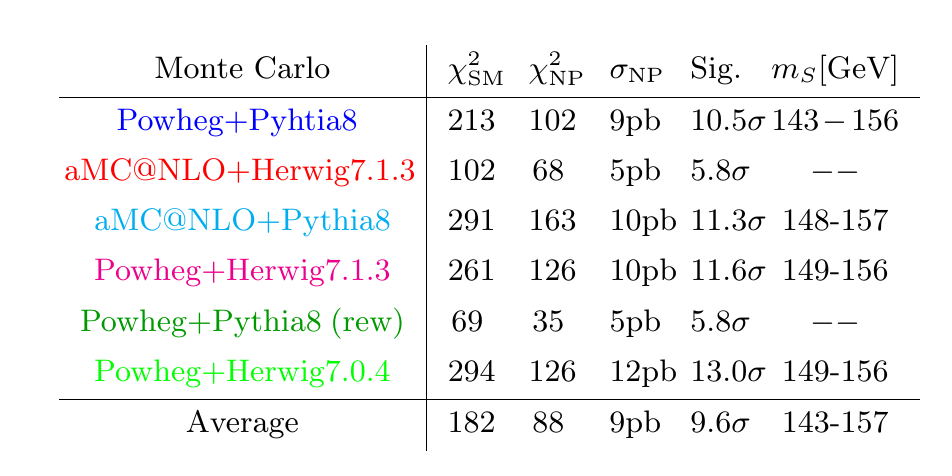}
   \caption{$\chi^2$ values, preferred cross-section ($\sigma_{\rm NP}$) significance (Sig.) etc. for the combined fit to $m^{e\mu}$ and $|\Delta\phi^{e\mu}|$ distributions for the six different SM simulations and their average. The $\chi^2_{\rm NP}$ is for our benchmark scenario with $m_S\approx152\,$GeV while the $m_S$ gives the preferred range of it from the fit.}
\label{tab:resmass150}
\end{figure}

ATLAS~\cite{ATLAS:2023gsl} considered six different SM simulations, as given in Table~\ref{tab:resmass150}, for the differential $t\bar t$ cross-section, including the invariant mass of the electron-muon pair ($m^{e\mu}$) and the angle between the leptons ($|\Delta\phi^{e\mu}|$). From the dashed lines in Fig.~\ref{fig:mlldata} one can see the poor agreement of the SM predictions with data.

In our simplified model, where the masses are fixed by the di-photon resonances to $m_S=152\,$GeV and $m_{S^\prime}=95\,$GeV, we assume that the dominant decay of $S$ is into pairs of $W$ boson, while $S^\prime$ decays dominantly to $b\bar b$. Note that this is naturally the case if $S$ is a SM-like Higgs and $S^\prime$ is the neutral component of an $SU(2)_L$ triplet with hypercharge~0.

We simulated in this setup the process $pp\to H\to S S^\prime\to (W W^{(*)}\to \ell^+\nu\ell^-\nu) b\bar b$, with $\ell=e,\mu,\tau$, using {\tt MadGraph5aMC@NLO}~\cite{Alwall:2014hca}, {\tt Pythia8.3}~\cite{Sjostrand:2014zea}, and {\tt Delphes}~\cite{deFavereau:2013fsa}. The results are shown as solid lines in Fig.~\ref{fig:mlldata}. One can see that for all SM simulations used by ATLAS, the agreement with data is significantly improved. In fact, as given in Table.~\ref{tab:resmass150}, the $\chi^2$ is reduced between $\Delta\chi^2=34$ and $\Delta\chi^2=158$, corresponding to a preference of the NP model over the SM hypothesis by $5.8\sigma$--$13.0\sigma$ while the average of the SM predictions results in $9.6\sigma$.

\begin{figure}[h]
\centering
    \includegraphics[width=0.8\linewidth]{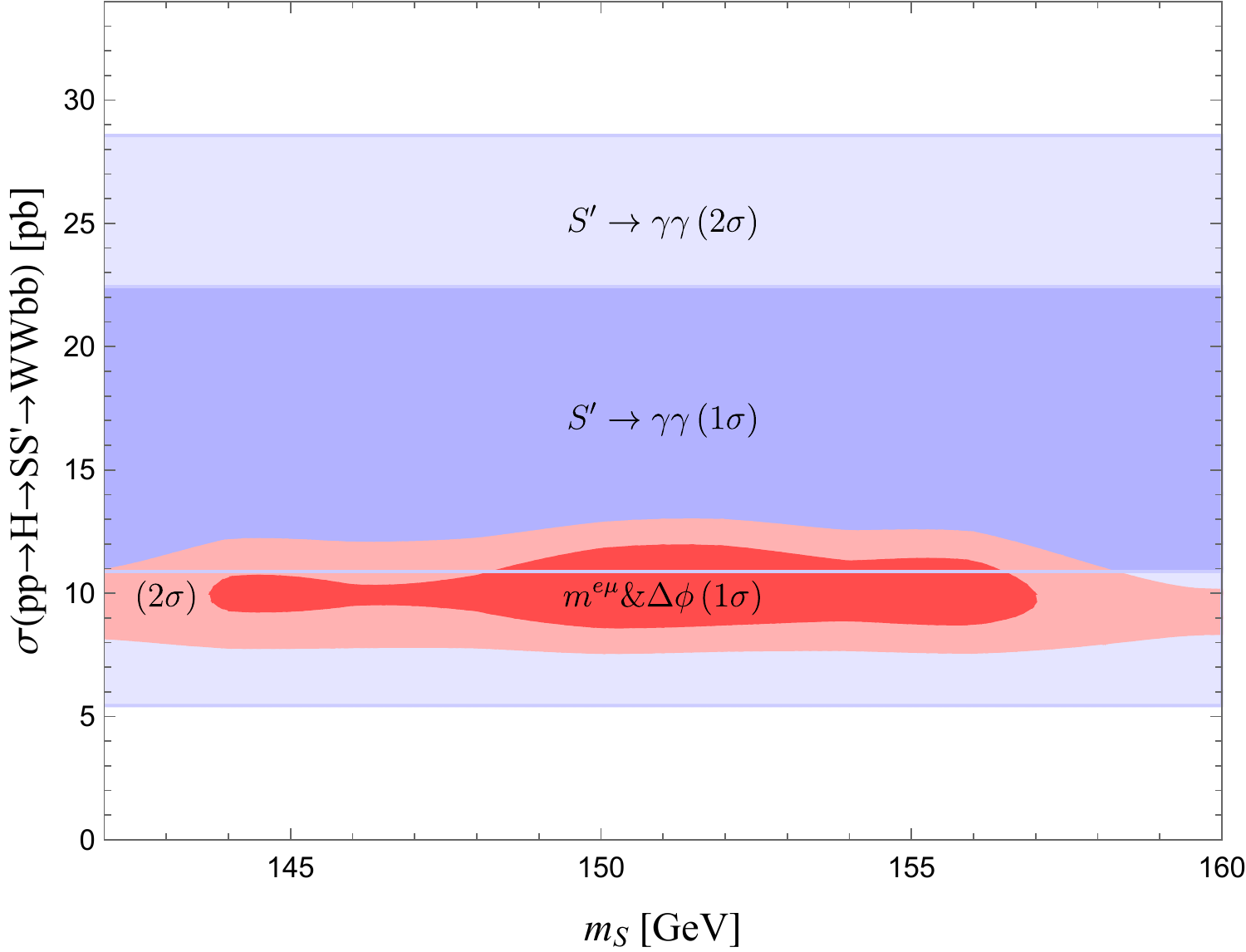}
    \caption{Preferred regions from the $t\bar t$ differential distributions (red) as a function of $m_S$ and the total cross section $pp\to H\to SS^\prime\to WWbb$ assuming $S^\prime$ to be SM-like and Br$[S\to WW]=100\%$. The blue region is preferred by the $95\,$GeV $\gamma\gamma$ signal strength.}
    \label{mSvsEpsNP}
\end{figure}

So far, we fixed the masses of $S$ and $S^\prime$ by using the hints for narrow di-photon resonances and took $m_H=270\,$GeV as a benchmark point which avoids constraints from SUSY and non-resonant di-Higgs searches. Let us now discuss the dependence on the masses. First of all, we checked that the effect of changing either the mass of $H$ or of ${S^\prime}$ is very small. This means that one can pick any value for $m_H$ so that this does not need to be counted as a degree of freedom in the statistical analysis. Now we can find the best-fit value in the $m_S$-$\epsilon_{\rm NP}$ plane by averaging the $x_i$ values of the six different MC simulations. For this, we generated 500k events each for $m_S$ between 140\,GeV and 160\,GeV in 2\,GeV steps and interpolated. As one can see from Fig.~\ref{mSvsEpsNP} the result is perfectly consistent with a mass of $152\,$GeV as suggested by the invariant mass of the di-photon system~\cite{ATLAS:2021jbf,Crivellin:2021ubm}.

Finally, let us consider the consistency of the preferred signal strength for $pp\to H\to SS^\prime$ with the di-photon excess at 95\,GeV. The branching ratio to photons of an SM-like Higgs with $95\,$GeV is $\approx\!1.4\times 10^{-3}$, Br$[S^\prime\to b\bar b]\approx\!0.86$~\cite{LHCHiggsCrossSectionWorkingGroup:2016ypw} and the production cross section at $13\,$TeV via gluon fusion is $\approx\!68$pb~\cite{LHCHiggsCrossSectionWorkingGroup:2016ypw}. Given that the preferred $\gamma\gamma$ signal at $95\,$GeV, with respect to a (hypothetical) SM-like Higgs with a mass of $95\,$GeV, is $\mu_{\gamma\gamma}=0.24^{+0.09}_{-0.08}$~\cite{Biekotter:2023oen} we can see in Fig.~\ref{mSvsEpsNP} that, under the assumption that $S^\prime$ is SM-like and Br$[S\to WW^*]\approx100\%$, the associated production of $S^\prime$, determined by the preferred signal strength from $t\bar t$, perfectly explains the excess.

\section{Conclusions}
\label{sec:conclusions}
ATLAS observed that the measured differential lepton distributions in its analysis of $t\bar{t}$ significantly differ from the predictions of the SM, which were obtained through various combinations of Monte Carlo simulators: ``No model (simulation) can describe all measured distributions within their uncertainties.'' This suggests that the measurement may be influenced by new physics contributions. 

We suggest that the $pp\to H\to SS^\prime \to WW b\bar b$ could constitute an NP background for the measurement. In fact, motivated by hints of di-photon resonances at around $95$GeV and $152$GeV, we have examined a benchmark scenario where $S^\prime$ and $S$ have corresponding masses and decay to $b\bar b$ and $WW$, respectively, with $m_H=270$GeV. This NP hypothesis is preferred over the SM hypothesis by at least $5.8\sigma$.

Furthermore, by varying $m_S$ we found that the preferred range is compatible with $m_S\approx 152\,$GeV, as motivated by the $\gamma\gamma$ and $WW$ excesses. Since the latter analysis employs a jet veto, this further disfavours the possibility that higher-order QCD corrections are the origin of the tension in the $t\bar t$ differential distributions. Assuming that $S^\prime$ is SM-like, this results in a di-photon signal strength in agreement with the $\gamma\gamma$ excess at $95$\,GeV. Since for $S$ a dominant decay to $W$ bosons is needed while the $ZZ$ rate should be low to respect the limits from inclusive $4\ell$ searches, this suggests that it could be the neutral component of the $SU(2)_L$ triplet with hypercharge 0.

\Acknowledgements
The work of A.C., S.B.~and G.C.~is supported by a professorship grant from the Swiss National Science Foundation (No.\ PP00P2\_211002). B.M.~gratefully acknowledges the South African Department of Science and Innovation through the SA-CERN program, the National Research Foundation, and the Research Office of the University of the Witwatersrand for various forms of support.

\bibliography{eprint}{}
\bibliographystyle{unsrt}
 
\end{document}